\documentstyle[prb,aps,twocolumn,epsfig]{revtex}
\begin{document}
\draft
%\tighten

\title{Absence of string order in the
anisotropic $S=2$ Heisenberg antiferromagnet}

\author{H. Aschauer and U. Schollw\"{o}ck}
\address{Sektion Physik, Ludwig-Maximilians-Universit\"{a}t M\"{u}nchen,
Theresienstr.\ 37, 80333 Munich, Germany}

\date{February 16, 1998}

\maketitle
\begin{abstract}
In this paper we study an antiferromagnetic 
Heisenberg $S=2$ quantum spin chain at $T=0$ with both interaction
and on-site anisotropy,
\[
H = \sum_{i} \frac{1}{2}(S^{+}_{i}S^{-}_{i+1}+S^{-}_{i}S^{+}_{i+1})
+J^{z}S^{z}_{i}S^{z}_{i+1}+D(S^{z}_{i})^{2}. 
\]
While the phase diagram of this chain for $S=1$ and
the isotropic point for $S=2$ are by now excellently understood, 
contradictory scenarios exist for the $S=2$ anisotropic phase diagram, 
which imply completely different mechanisms of the emergence of the classical
$S\rightarrow\infty$ limit, hence the importance of the question.
One main scenario predicts the emergence of a cascade of phase transitions
not seen in the $S=1$ case, based on an analysis using the 
Affleck-Kennedy-Lieb-Tasaki model. Another scenario is in favor of an
almost classical phase diagram for $S=2$, from which the $S=\infty$
phase diagram evolves quite trivially; the $S=1$ case then emerges
as very special with its dominant quantum effects. Numerical studies
have so far not been conclusive.

Using the Density Matrix Renormalization Group, we study gaps and 
correlation functions in the anisotropic $S=2$ chain. The
question of the existence of hidden topological order in the 
anisotropic $S=2$ chain is emphasised, as it distinguishes between the
proposed scenarios. Careful extrapolation allows us to show that
the topological order is zero in the thermodynamical limit 
in all disordered phases, in particular
in the new phase interposed between the Haldane and large-$D$ phases.
The decay of the hidden order is substantially faster than that of
$S^{+}S^{-}$ correlations and excellently fitted by power-laws
typical of massless phases.
This excludes the AKLT-model based scenario in favor of an almost
classical phase diagram for the $S=2$ and thus higher spin chains.
 
\end{abstract}
\pacs{75.50.Ee, 75.10.Jm, 75.40.Mg}

\narrowtext

\section{Introduction}
Interest in quantum spin chains as examples of low-dimensional 
strongly correlated quantum systems was greatly revived in 1983 by
Haldane's famous conjecture\cite{Haldane 83} 
of a fundamental difference between
half-integer and integer quantum spin chains, a phenomenon without
equivalent in higher dimensional systems. 

In the meantime, Haldane's conjecture has been firmly established,
while not rigorously proven. The numerical values of the 
predicted gaps and finite correlation
lengths for isotropic integer spin chains are by now extremely well
known both in the 
$S=1$\cite{Botet 83,White 92,White 93a,Golinelli 94,gaps1} and the 
$S=2$\cite{Schollwoeck 95,Schollwoeck 96,Yamamoto 95,gaps2} cases, which constitute the
extreme quantum limit. Furthermore, for both cases Haldane's predictions
of a massive-relativistic dispersion relation have been 
confirmed\cite{White 93a,Yamamoto 95,Golinelli 92b}.

Haldane's conjecture was derived from a non-linear quantum sigma model,
which is an effective low-energy continuum theory of the isotropic
Heisenberg model in the semiclassical limit $S\rightarrow\infty$. It
does not provide immediate insight in the ground state of integer spin chains. 

The Affleck-Kennedy-Lieb-Tasaki (AKLT) model\cite{Affleck 87} provides 
a profound physical understanding of gapped spin liquids. The AKLT model
is the Hamiltonian which gives rise to the following exact
ground state: each spin of length $S$ is decomposed into $2S$ completely
symmetrized $S=\frac{1}{2}$ spins. Pairs of such spins on neighboring
sites are linked up by singlet bonds, giving rise to $S$ singlet bonds
between sites. In the case of $S=1$, all important features of the
isotropic Heisenberg model and the AKLT model,
which has the Hamiltonian
\begin{equation}
H_{AKLT}^{S=1} = \sum_{i} {\bf S}_{i}{\bf S}_{i+1} + 
\frac{1}{3}({\bf S}_{i}{\bf S}_{i+1})^2
\end{equation} 
are in qualitative 
agreement\cite{White 93a,Girvin 89,Kennedy 90,Kennedy 92,Elstner 94}. 

A particularly striking feature of both models is the existence
of a hidden non-local topological order, also known as string 
order\cite{Girvin 89,Kennedy 92,Nijs 89,Oshikawa 92},
\begin{equation}
O(i,j) = \langle S^z_i e^{i\pi\sum_{k=i+1}^j S^z_k} S^z_j \rangle,
\end{equation}
which does not vanish 
in the limit $|i-j|\rightarrow\infty$. For the AKLT model, the value is
$\frac{4}{9}$, for the Heisenberg model $0.37432$\cite{White 93a,Girvin 89}. 
In the case of
a $S=1$ chain, it describes a
hidden antiferromagnetic order of $S^z_i=\pm 1$ spin states, which is diluted
by arbitrary numbers of interposed $S^z_i=0$ spin states and thus
not seen in the antiferromagnetic order parameter.

The existence of the hidden order can further be associated with the
breaking of a $Z_2 \times Z_2$ symmetry in the Haldane 
phase\cite{Kennedy 92}, as well 
as with the existence of a four-fold degenerate ground state of an open
Heisenberg chain in the Haldane phase\cite{Kennedy 90}. 
For $S=1$, the success of the AKLT-model can be understood as it is a 
quantum disorder point, a remnant of a classical phase 
transition\cite{Schollwoeck 96b}.
For higher spins, as we will see,
both the existence of hidden order and its signification are not clear.

In this paper, we want to make progress towards understanding how the
classical phase diagram emerges out of the substantially different
$S=1$ phase diagram. Examining different 
scenarios\cite{Schollwoeck 95,Oshikawa 92,Schulz 86,Khveshchenko 87,Oshikawa 95}, 
we will see that
their fundamental difference regards whether or not there exists hidden
order in the $S=2$ phase diagram for suitable anisotropies.
Our results will allow to decide quite clearly in terms of a scenario
without hidden order for $S=2$. We finish by addressing briefly why
the AKLT model is not successful in explaining the anisotropic
$S=2$ phase diagram.

\section{$S=1$ and $S=\infty$ Phase Diagrams}

In real physical systems, such as NENP as the quintessential $S=1$
spin chain\cite{Renard 87}, 
anisotropies are always present, in the case of NENP an
easy-plane on-site anisotropy $D\approx 0.18$\cite{Golinelli 93}. 
In order to link experiments to
theory, it is therefore necessary to understand the phase diagram
with anisotropies. At the same time, this may provide deeper theoretical
insight into the interplay of the different phases.

\begin{figure}
\centering\epsfig{file=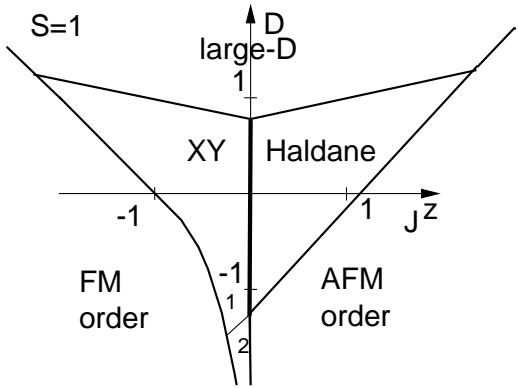,scale=0.85}
\vspace{0.3truecm}
\caption{$T=0$ phase diagram of the $S=1$ anisotropic Heisenberg
antiferromagnet. Transition lines are schematic; the XY-Haldane
transition line is at $J^z=0$ exactly; the FM-XY transition line
includes $D=0$, $J^z=-1$. For $J^z=1$, the Haldane-large-$D$ transition
is at $D=0.99(1)$; for $D=0$. the Haldane-AFM transition at $J^z\approx
1.2$. The XY phase actually comprises two different XY phases 1 and 2.}
\label{fig:s1diag}
\end{figure}

{\em $S=1$ phase diagram.} The $S=1$ phase diagram, 
as the physically most relevant one with at the
same time the strongest quantum effects for an integer spin chain,
has been studied extensively and is by now extremely well 
understood\cite{Botet 83,Nijs 89,Schulz 86,Sakai 90,Schulz 86b,Golinelli
  92b,Yajima 94,Kitazawa 96}.
Let us summarize the main features (Figure \ref{fig:s1diag}).  

Both for substantially strong $J^z$ anisotropy and sufficiently small $D$,
one finds Ising-like ordered (anti)ferromagnetic phases,
depending on the sign of $J^z$. The transition lines are not very far
from those in the classical limit (see below).

All other phases are disordered, but otherwise physically quite distinct:
For large $D$, all $S^z_{i}\neq 0$ states are suppressed and a singlet phase
emerges, with a gap proportional to $D$, as the gap is essentially the energy 
to excite
one spin to $S^{z}_{i}=\pm 1$\cite{Papanicolaou 89}. 
Below that, there are two phases. Their separation line is
strictly $J^z=0$, as first shown by den Nijs and Rommelse\cite{Nijs 89}, 
and later on
confirmed numerically with increasing 
precision\cite{Botet 83,Sakai 90,Schulz 86b,Yajima 94,Kitazawa 96}. 
For $J^z>0$, there is
a gapped quantum liquid, the Haldane phase, including the isotropic
Heisenberg model, with the properties described above. For $J^z<0$, there
is an XY-phase with power-law decay in the $\langle S^{+}_i S^-_j \rangle$
correlations. In fact, the XY phase decomposes into two phases with
different correlation functions, linked by an Ising-like 
transition\cite{Nijs 89,Schulz 86}.
Phase transitions out of FM order are first order, from AFM order to
the large-$D$ phase first order, from Haldane to XY, and from XY to large-$D$
Kosterlitz-Thouless like, and the Haldane to large-$D$ transition has been
likened to a preroughening transition in RSOS models\cite{Nijs 89}.

\begin{figure}
\centering\epsfig{file=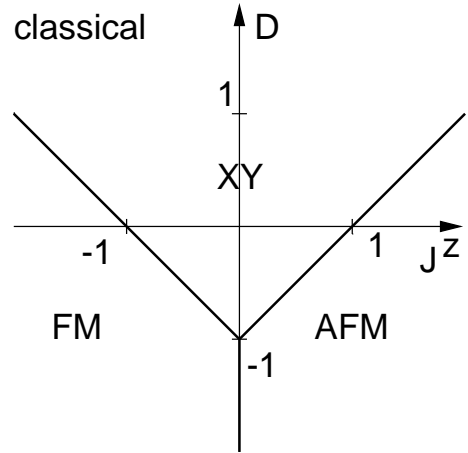,scale=1.0}
\caption{$T=0$ phase diagram of the classical anisotropic Heisenberg
antiferromagnet. All transition lines are exact.}
\label{fig:classical}
\end{figure}

{\em Classical phase diagram.} On the other hand,
the $T=0$ phase diagram in the semiclassical limit $S\rightarrow\infty$ 
is very easily obtained (Figure \ref{fig:classical}). 
The quantum spins of length $S$ are replaced by
unit vector spins which do commute. This is in order as all terms in the
Hamiltonian are of the order $S^{2}$, such that this factor can be scaled
out. One finds three phases:

\begin{enumerate}
\item $J^{z}>0$ and $D<J^{z}-1$: ordered antiferromagnetic Ising-phase 
(AFM)
\item $J^{z}<0$ and $D<-J^{z}-1$: ordered ferromagnetic Ising-phase 
(FM)
\item $D>|J^{z}|-1$: ordered antiferromagnetic XY phase (XY):
all spins in-plane
\end{enumerate}

\section{Emergence of the Classical Limit}

How does this classical phase diagram emerge from the $S=1$ phase diagram?
In recent years, various scenarios have been proposed.

\begin{figure}
\centering\epsfig{file=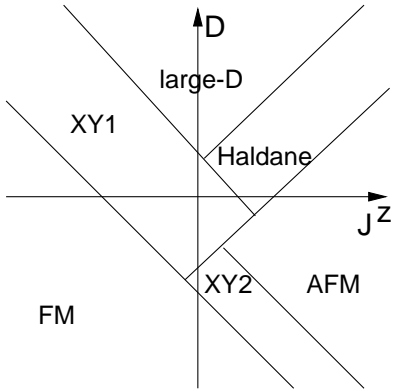,scale=0.85}
\vspace{0.3truecm}
\caption{Phase diagram for $S>1$ as predicted by bosonization. The figure
is not to scale. Adapted from Ref.\ 20.}
\label{fig:boson}
\end{figure}

1. Bosonisation\cite{Schulz 86} 
predicts for all integer spins $S$ essentially the same 
scenario as for $S=1$ (Figure \ref{fig:boson}). Bosonisation can be expected
to predict the nature of the phases found and their topology (i.e.\ which
phases are joined by a transition), but not to yield quantitative values
for the phase boundaries. 

For $S=1$, bosonisation indeed predicts all phases and the correct
topology of the phase diagram; it merely misses that the large-$D$, AFM and
FM phases meet respectively for very large anisotropies, beyond the
limit of applicability of bosonisation. For higher spins, it is predicted 
in particular that for an isotropic
exchange interaction $J^{z}=1$
there is exactly one transition from the Haldane phase
to the large-$D$ phase for $D\rightarrow\infty$. From other 
arguments\cite{Papanicolaou 89}
it can be concluded that for large $S$ the large-$D$ phase will
pushed out to $D\rightarrow\infty$, whereas the gap of the Haldane phase
vanishes; in the $S\rightarrow\infty$ limit the gapless XY-phase emerges
(but for the isotropic point, any $D>0$ will then favour in-plane spins
for $J^z=1$).

Numerical studies\cite{Schollwoeck 95,Oshikawa 95,Nomura 97} 
are by now in excellent agreement that already for the
$S=2$ case there are not only two, but three phases when taking $D\rightarrow
\infty$ for positive $D$ and $J^{z}=1$. While the results of
Ref.\ \onlinecite{Oshikawa 95} are not reconcileable with bosonisation,
those of Refs.\ \onlinecite{Schollwoeck 95} and \onlinecite{Nomura 97}
could be interpreted as being in accordance with bosonization, if one
allows for a major deformation of the bosonization phase diagram.
In the scenario we are going to show numerically below, bosonisation will be
found to predict the correct phases, but we will also see that the necessary
deformation of the bosonisation phase diagram changes it almost beyond
recognition.

\begin{figure}
\centering\epsfig{file=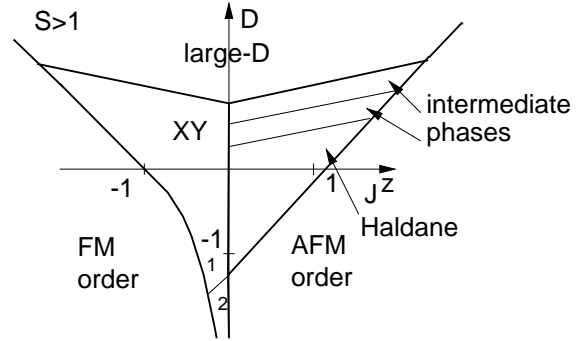,scale=0.75}
\vspace{0.3truecm}
\caption{$T=0$ phase diagram for $S>1$ anisotropic Heisenberg
antiferromagnets proposed by Oshikawa\cite{Oshikawa 92}. 
Between the Haldane and
large-$D$ phases, $S-1$ intermediate phases appear, characterised by
an alternatively finite and zero string order.} 
\label{fig:s2oshi}
\end{figure}

2. A second scenario was proposed by Oshikawa\cite{Oshikawa 92} 
in the spirit of the 
Affleck-Kennedy-Lieb-Tasaki (AKLT) model (Figure \ref{fig:s2oshi}). 

\begin{figure}
\centering\epsfig{file=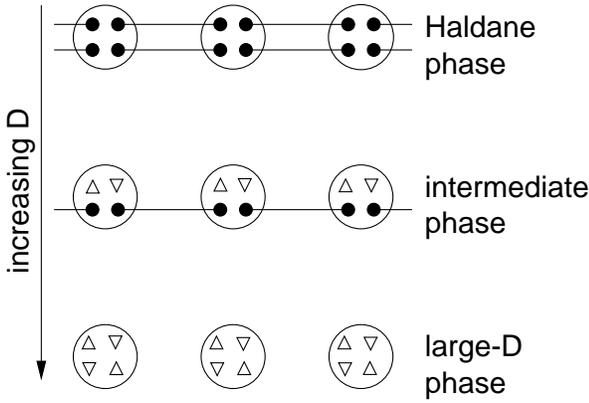,scale=0.85}
\vspace{0.3truecm}
\caption{Mechanism proposed by Oshikawa for the emergence of the
large-$D$ phase out of the Haldane phase for $J^z>0$ with increasing
$D$. Circles represent spins of length $S$ (here 2), 
dots completely symmetrized
spins-$\frac{1}{2}$. Bonds are singlet bonds. Up and down triangles
symbolize symmetrized spins which contribute zero magnetisation and
are not bonded.}
\label{fig:oshiprop}
\end{figure}

Oshikawa\cite{Oshikawa 92} has shown that the hidden topological order
parameter (the string order parameter) disappears exactly in the
AKLT model, describing the isotropic chain, for all
even integer spins $S=2,4,\ldots$, while it yields a finite value for
odd integer spins $S=1,3,\ldots$. On the other hand, Oshikawa argues
that the effect (assuming for the moment an isotropic exchange interaction,
$J^{z}=1$) of an increase in 
the easy plane anisotropy $D>0$ is to successively suppress
the large $S^{z}$ states: first $\pm S$, then $\pm (S-1)$, and so on.
This yields in his analysis a succession of spin chains 
with effective spin lengths $S, S-1, \ldots, 1, 0$.
The enviseaged succession of spin models can be schematically drawn in the
AKLT picture as in Figure \ref{fig:oshiprop}. 

As the string order parameter should be alternatingly zero and non-zero
in the even and odd integer spin chains, Oshikawa predicts a succession
of $S+1$ phases with $S$ phase transitions, beginning with the isotropic
Haldane phase (spin length $S$) and ending with the large-$D$ phase,
assimilated to a spin length 0. The transitions are characterized by the
(dis)appearance of a finite string order parameter.
The classical limit would then emerge by a softening of the quantum phase
transitions and a disappearance of possible gaps in the $S+1$ phases. 

The AKLT scenario works excellently for the $S=1$ case; and it is extremely
well established that a similar scenario, also proposed by 
Oshikawa\cite{Oshikawa 92}, of
multiple phase transitions works extremely well to describe successive
phase transitions in an isotropic Heisenberg integer spin antiferromagnet
with increasingly strong dimerization of the interaction bonds, in excellent
agreement with field theoretical predictions\cite{Affleck 87b} and
numerical verification\cite{Kitazawa 96}. All this makes this a very
strong scenario.

The existence of an intermediate phase, as should exist for $S=2$,
was first established in Ref.\ 
\onlinecite{Schollwoeck 95}, and later confirmed in Refs.\
\onlinecite{Oshikawa 95} and \onlinecite{Nomura 97}. 
Oshikawa et al.\cite{Oshikawa 95}, using Quantum Monte Carlo 
on systems up to $L=160$,
claim to see, at least
for $D\approx 1.2$ and $J^z=1$, 
a non-vanishing string order parameter, which they
extrapolate to a thermodynamic limit of $1 \times 10^{-3}$ at that
particular point. They interpret furthermore the behavior of spin-spin 
correlations as indicative of a phase transition for $J^{z}\approx 0$.
Furthermore, a sign-alternation exists in the string order at the
isotropic point, but not at $D=1.2$.
These results would support Oshikawa's scenario of multiple transitions.

\begin{figure}
\centering\epsfig{file=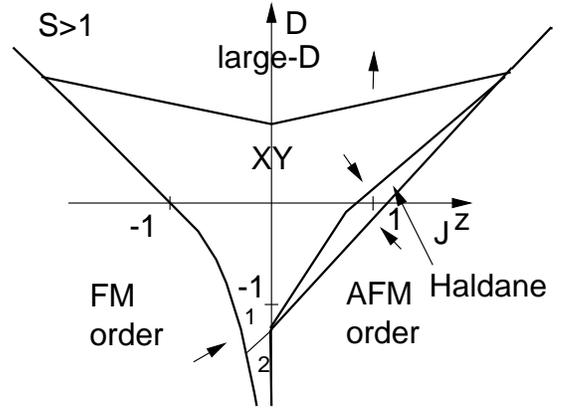,scale=0.9}
\vskip 1truecm
\caption{$T=0$ phase diagram for $S\geq 2$ anisotropic Heisenberg
antiferromagnets proposed by Schollw\"{o}ck et al. The Haldane phase
is almost squeezed out, the large-$D$ phase moves up to infinity for
large $S$. The XY phases extend rapidly with $S$, $S=2$ is almost
classical already. The transition line to large-$D$ is not to scale
(shown too low); the Haldane phase is exaggerated.}
\label{fig:s2scholl}
\end{figure}

3. A third scenario results from numerical work by Schollw\"{o}ck et 
al\cite{Schollwoeck 95} (Figure \ref{fig:s2scholl}). It actually has
a precursor in the works of Khveshchenko and Chubukov\cite{Khveshchenko 87}. 
They argue that the large-$D$ phase
moves as $S^{2}$ up to 
$D\rightarrow\infty$ for $S\rightarrow\infty$, while the Haldane
phase recedes to the isotropic point. The XY phase already present in the
$S=1$ phase diagram invades the emptied part of the phase diagram; in the
XY phase the hidden order parameter is zero in the thermodynamic limit.
The $S=1$ case with its strict $J^{z}=0$ transition line between XY and
Haldane phase 
emerges thus as particular. 
Numerically it is found that the Haldane phase is reduced to a small
stripe next to the antiferromagnetic phase and squeezed out for $S\rightarrow
\infty$, as the gap is suppressed exponentially as $e^{-\pi S}$; 
the large-$D$ phase moves up to $D\rightarrow\infty$, and the
classical phase diagram emerges in a particularly simple fashion.
There is no clear information on the behavior of the Haldane phase at its
two end points. It is not clear whether (i) the Haldane phase touches
the large-$D$ phase at its top right end point and (ii) whether XY and AFM
phase meet at its left end. If both are true, then not only the same phases,
but also the same phase boundaries as in the bosonization picture emerge;
the phase diagram looks still quite different. 

Numerically, the
existence of an additional phase between the Haldane and the large-$D$
phase has been demonstrated\cite{Schollwoeck 95} 
using the DMRG\cite{White 92,White 93b}.
For $J^{z}=1$, the transition points were placed at $D_{c_{1}}=0.04(1)$
and $D_{c_{2}}=3.0(1)$. The intermediate phase was identified as gapless,
with exponentially decaying $\langle S^{z}_{i}S^{z}_{j}\rangle$ correlations,
power-law like decay in the $\langle S^{+}_{i}S^{-}_{j}\rangle$ correlations,
and a vanishing string order parameter, in support of the above picture.
However, available computational
ressources did not allow a very precise determination of the quantities
involved (lengths up to 200, keeping 200 block states in the DMRG).

Recently, Nomura and Kitazawa, combining exact diagonalization and
field theory techniques, have shown\cite{Nomura 97} that there
are, on the $J^{z}=1$ line, two phase transitions, one from the Haldane
phase to an XY phase, one from an XY phase to a large-$D$ phase. They
localise their transitions at  $D_{c_{1}}=0.043$
and $D_{c_{2}}=2.39$. While the first result is in good agreement with
DMRG results, the second is not. As the gap opens exponentially weak,  
the DMRG, which determines the gap directly, has difficulty
in extracting such a transition point; fits to an exponential
law are hindered by the presence of small additive errors which become
important when the gap approaches $\approx 0.001$. However, evaluating new, 
more precise DMRG data with
this fit formula shows that $D_{c_{2}}$ was overestimated in Ref.\ 
\onlinecite{Schollwoeck 95}; with higher precision, we can show a
gap down to $D\approx 2.5$. 
In essence, the findings do agree,
and the identification of the phase is the same in both works, at least
on the $J^{z}=1$ line. 
The numerical data in Ref.\ \onlinecite{Nomura 97} 
leaves however open the possibility that the phase
with a non-vanishing string order parameter proposed by Oshikawa might
be located closer to the point where the large-$D$ and antiferromagnetic
phase meet directly. We will address this question now. 

\section{String Order in the $S=2$ Phase Diagram}

1. Before we discuss our results on gaps and correlation functions,
a remark on the calculation of the string order parameter is in order,
as the crucial information is whether it is finite or not in the thermodynamic
limit.
Conventionally, the string order parameter is defined by
\begin{equation}
O(i,j) = \langle S^{z}_{i} e^{i\pi\sum_{k=i+1}^{j} S^{z}_{k}} 
S^{z}_{j} \rangle .
\end{equation}
This means that the exponential string includes {\em one} of the two
spins correlated. In the $S=1$ literature, numerically often two
alternative quantities are calculated: an order parameter with a string
including {\em both} correlated spins,
\begin{equation}
O^{2}(i,j) = \langle S^{z}_{i} e^{i\pi\sum_{k=i}^{j} S^{z}_{k}} S^{z}_{j} \rangle ,
\end{equation}
and an order parameter including {\em none} of the correlated spins,
\begin{equation}
O^{0}(i,j) = \langle S^{z}_{i} e^{i\pi\sum_{k=i+1}^{j-1} S^{z}_{k}} 
S^{z}_{j} \rangle .
\end{equation}
Obviously, this changes some of the signs of contributions to the expectation
value, depending of the actual value of the correlated spins. In the $S=1$
case, it is obvious that a sign change occurs exactly in all contributions
that are not zero, where both $S^{z}_{i,j}\neq 0$. Therefore
\begin{equation}
-O^{2}(i,j)=O(i,j)=-O^{0}(i,j)
\end{equation}
strictly. In the $S=2$ case, this is no longer true: $S^{z}=\pm 2$ and
$S^{z}=\pm 1$ lead to different sign behavior in the three definitions.
It is however, if computational ressources are stretched to their limit,
more efficient for memory consumption reasons to calculate either $O^{0}$
or $O^{2}$ in the DMRG method because then the string can be decomposed into
two symmetric parts, if $i$ and $j$ are chosen symmetric about the chain
center of an open chain. Only one needs to be stored then. 

We have checked in lower precision calculations (200 states kept, length 200) 
that in the region of
importance the following inequality is always observed:
\begin{equation}
|O^{2}(i,j)| \leq |O(i,j)| \leq |O^{0}(i,j)|. 
\end{equation}
The very small difference in these correlations decreases with increasing $|i-j|$ both
absolutely and relatively. Therefore, in the following, we always refer to 
the calculation of $O^{0}(i,j)$. It provides both an upper bound to
$O(i,j)$ and converges
to the same thermodynamic limit for $i-j\rightarrow\infty$. 
In fact, on all our plots the difference
between the various correlations would be invisible to the eye.

2. An observation that can be made numerically is that the string order
parameter at $D=0$, $J^{z}=1$ decays with alternating sign, whereas in
the intermediate phase it decays with no sign-alternation. This might
be attributed to a substantial difference between the various $S=2$
phases, as it has been done in Ref.\ \onlinecite{Oshikawa 95}. 
Alternatively, this might naively be attributed to the fact that
$D>0$ suppresses both large $S^{z}$ values and leads to a very fast decay
of $S^{z}S^{z}$ correlations: With $J^{z}=1$, spin-spin correlations
are antiferromagnetic for arbitrary $D$. This introduces a fundamental
sign-alternation in the string order. What is effectively observed, now
depends on the string:
For $D\approx 0$ and $J^{z}=1$, the close
neighbourhood of the Ising AFM phase indicates that $S^{z}=\pm 2$ states
dominate in the string, contributing factors 1. The string order parameter
therefore exhibits sign-alternation. When $D$ is successively increased,
$S^{z}=\pm 1$ and then $S^{z}=0$ states dominate, introducing a sign
alternation in the string for the former and no alternation for the latter
case, such that the total order parameter is non-alternating and then
alternating again.

Calculating chains up to length $L=200$ with the DMRG, 
we find for $D\geq 0.02$ that
the string order parameter shows sign-alternation for small distances,
but no sign-alternation for longer distances. As we find (see below) a
rather strong dependence of the string order parameter on the total
system size if the string length becomes of the order of the system
size, we have localized this ``crossover'' of the sign-behavior
for various total system lengths.
For $D\geq 0.03$, the crossover point is sufficiently small compared to the
total system length that it may be considered almost converged; the
numerical data does not exclude the possibility that the sign crossover
can already be observed for smaller $D<0.02$. We find crossover points as
in Table \ref{tab:table1}.

The sign crossover can be attributed to rather strong short ranged
Neel like order, which is still present for small $D$, but does not
survive for larger distances. It is of no importance whatsoever for
larger values of $D$. The change in behavior is gradual and 
not directly linked 
to the critical value of $D=0.043$

For the large-$D$ phase, the second change in sign behavior is not observed 
up to $D=3.5$, which is already deep in the large-$D$ phase (gap of the order
0.05).

This establishes that there is no direct relationship between the
phase transitions and the sign behaviour of the string order; a
non-alternating decay is no unique feature of the intermediate phase. 

3. Furthermore, we have compared the decay of various correlation functions
in the XY phase. In all points, we find a very fast exponential decay of
$S^{z}S^{z}$ correlations, and power-law decays in the string order
and the $S^{+}S^{-}$ correlations. The decay of the string order is much
faster than that of the typical XY phase correlations,
even though the difference decreases towards the upper right corner
of the XY phase. We see no change
in behavior between $J^{z}<0$ and $J^{z}>0$. To give an example,
decays follow laws $x^{-\alpha}$, with $\alpha=0.13$ for $J^z=-0.5$,
$D=1.0$ and $\alpha=0.18$ for $J^z=0.5$, $D=1.0$ for the $S^{+}S^{-}$
correlations and  $\alpha=2.11$ for $J^z=-0.5$,
$D=1.0$ and $\alpha=1.47$ for $J^z=0.5$, $D=1.0$ for the string order. 
This is indicative of the XY nature of the intermediate phase.

\begin{figure}
\centering\epsfig{file=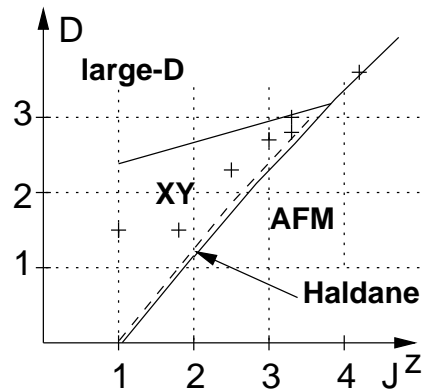,scale=1.0}
\vspace{0.3truecm}
\caption{Location of points which were calculated for chain lengths
$L=600$, keeping 400 states for the DMRG. 
The transition lines were obtained numerically; the AFM
transition line is determined up to $D=D_c \pm 0.05$ or better;
the Haldane phase has a width of $\approx 0.05$ or smaller and is thus
hardly distinguishable from the XY phase, also due to the small gap.
The $D$ transition line is possibly slightly too high, as DMRG has 
difficulty in seeing extremely small gaps.}
\label{fig:points}
\end{figure}

4. Before we discuss results on the string order in the intermediate phase,
let us first try to locate it with more precision. We consider only the
part of the phase diagram with $J^{z}>0$ and $D>0$ (Figure \ref{fig:points}).

The transition line from the AFM phase to the Haldane or large-$D$ phases
can be easily located with the behavior of the $S^{z}S^{z}$ correlation
function. Considering chains up to length $L=300$, this transition line
was determined with a precision of $\pm 0.05$ in the critical $D$-value.

The Haldane phase is very narrow (cf.\ former results for 
$J^{z}=1$\cite{Schollwoeck 95,Nomura 97}), approximately 0.1 wide in units of
$D$. A more precise determination has not yet been carried out, only
at $J^z=1.0$ (considered above) and $J^z=3.3$ (see below).

\begin{figure}
\centering\epsfig{file=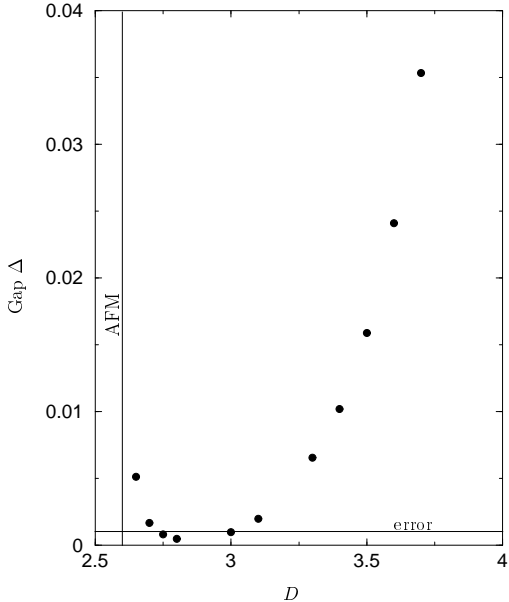,scale=0.5}
\caption{Location of extrapolated gaps $\Delta$ for $J^z=3.3$ vs.\ anisotropy
$D$. For $D=2.6$, the chain is in the AFM phase. The bottom line
$\Delta=0.001$ indicates our estimate of the precision limit. For smaller 
gaps, it cannot be clearly decided whether they are finite. The gapped phase
on the left is the Haldane phase, the gapped phase on the right the large-$D$
phase. Note that the gap curve would be almost flat, if the same scale
were used for $D$ and $\Delta$.}
\label{fig:gaps}
\end{figure}

The crucial question is the location of the transition from the
intermediate to the large-$D$ phase, because the gap in the
large-$D$ phase opens exponentially slow. 
Here, we have calculated the
gap in high-precision calculations (300 states kept, length 300) for
various values of $D$ for $J^{z}=1; 2.5; 3.3$. The fact that the 
gap is very small (or vanishes) implies an extremely large or divergent $\xi$,
which has the advantage that a simple $L^{-1}$-extrapolation of the
gap should provide a rather precise estimate of the gap
(cf.\ the detailed discussion of such extrapolations in Refs.\
\onlinecite{Schollwoeck 95,Schollwoeck 96}). DMRG
errors cancel to some extent as both ground and first excited state
have similar truncation errors (approximately $10^{-8}$ or better for $L=600$ 
and
$M=400$ states kept). Variation of the number $M$ of kept states
reveals that we overestimate gaps slightly. Extrapolation in $M$ as well
as gap calculations for points where the gap is known to disappear let
us estimate that gap values $\Delta>0.001$ are really finite. 
Consider
Figure \ref{fig:gaps}. For $J^z=3.3$, the truncation error is comparatively
large, $\approx 2 \times 10^{-8}$. The very small gaps observed leave two
scenarios: (i) between the Haldane phase ($D\approx 2.65$) and the
large-$D$ phase (for D larger than $D\approx 3.0$) there is a small
intermediate phase; 
(ii) Haldane and large-$D$ phase meet directly, at $D\approx 2.8$.
For smaller $J^z$, the two phases can be clearly separated, still there is
some uncertainty as regards the transition to the large-$D$ phase. 
This leads us to the diagram shown in Figure \ref{fig:points}; the upper
transition line might in reality be somewhat lower (by no more than 0.2
units in $D$ close to the upper right end of the XY phase; the transition
point for $J^z=1$ is essentially exact). In any case, if we 
{\em over}estimate the
size of the intermediate phase, there is no finite string order in the true
intermediate phase if we don't find it in our oversized one. So our argument
is not weakened. In fact, most of our points are far from the transition line
anyways.

5. Within the intermediate phase thus identified
we are now going to calculate the thermodynamic limit of the string order
for a large number of points. 

To discuss the possibility of an Oshikawa-like phase in the
upper right corner, we will systematically investigate points for
large $J^{z}$. 

Our procedure to calculate the thermodynamic limit of the string order 
parameter is as follows. Using the DMRG, we calculate for a given set of
$(J^{z},D)$ the string order parameter
\begin{equation}
O(N,L,M) = \langle S^z_i e^{i\pi\sum_{k=i+1}^{j-1}} S^z_j \rangle
\end{equation}
with $N=|i-j|$, $i$, $j$ symmetric about the center of an open chain of length
$L$ studied using a reduced Hilbert space of $M$ states per block. 
 
To obtain the 
thermodynamic limit $N\rightarrow\infty$, we first have to eliminate the DMRG
typical dependence on $M$ of an order parameter for given $N$, $L$. This
extrapolation can be done by fitting the string order against the DMRG
truncation error. We find that for $M=400$ the errors
in the extrapolated results are of no relevance to the following 
extrapolations.

\begin{figure}
\centering\epsfig{file=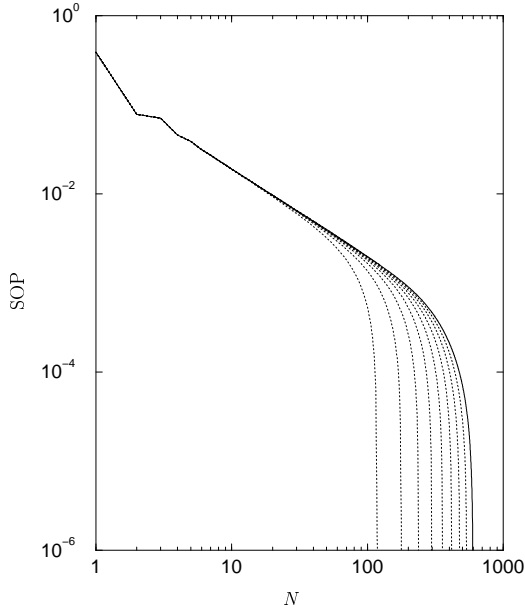,scale=0.5}
\caption{Double-logarithmic plot of the string order $O(N,L)$ for various
system lengths $L=120$, 180, 240, 300, 360, 420, 480, 540, 600 for
$J^z=2.5$, $D=2.3$, with $M=400$ DMRG block states kept. The effect of the
finite chain length as well as the convergence to a power law over several
orders of magnitude are clearly visible.}
\label{fig:loglog}
\end{figure}

Having obtained $O(N,L)= \lim_{M\rightarrow\infty} O(N,L,M)$, we now have
to take the thermodynamic limit first in $L$ and $N$. 
The numerical result is an interplay of both the finite $N$ and $L$; as can be
seen from the example shown in 
Figure \ref{fig:loglog}, the influence of finite $L$ is
strong. In Ref.\ \onlinecite{Oshikawa 95}, these two limits are
essentially taken at the same time, with $N=L/2$; to take this into account,
a conformal field theory formula for correlations in systems with
periodic boundary condition was used. As the power-law extracted seems in
very good accordance with our findings, which consider however much larger
systems ($L=600$ vs.\ $L=160$), we attribute the deviations found in that
reference to the evaluation of the Monte-Carlo data used.
Keeping $N$ fixed, we
find that the convergence of the string order parameter to thermodynamic
limit value for fixed $N$ is extremely well fitted by a $L^{-1}$ law, if
$N$ is up to $\approx 0.6 L$; for larger $N$,
effects from the open chain ends start to dominate. 
For $N$ less than $\approx 0.4L$, the
extrapolation correction is virtually negligible.

We thus get $O(N) = \lim_{L\rightarrow\infty} O(N,L)$. This value can now
safely be extrapolated to $N\rightarrow\infty$. We find numerically that
$O(N)$ vs.\ $N$ can for $N$ not too small (such that lattice effects
disappear) be extremely well fitted by power laws such as
\begin{equation}
O(N) = C(J^z,D) N^{-\alpha(J^z,D)},
\end{equation}
as expected in a critical phase, where $\alpha$ varies slowly. The exponent
$\alpha$ is given in Table \ref{tab:table2} for our representative highest precision points; all points calculated at less computational expense reflect
the given results. The law expected in a massless phase
from conformal field theory is 
\begin{equation}
\langle \phi(x)\phi(y) \rangle = \frac{C}{|x-y|^{2h}}
\end{equation}
for conformal weight $h=\overline{h}$. We observe therefore a decay of
the string order to zero in the thermodynamic limit in the intermediate
phase, following a typical decay law of a massless phase, which we identify
as an XY phase. It can also be observed that the exponent $\alpha$ decreases
with increasing $J^z$ and increases with increasing $D$.

\section{Conclusion}

In our opinion, our numerical findings together with the analysis of
Nomura and Kitazawa\cite{Nomura 97} show that the emergence of the 
classical limit is governed by the third scenario of an almost
classical phase diagram for $S=2$ (Figure \ref{fig:s2scholl}). 

Why the failure of the AKLT model to account for the phase diagram of the
$S=2$ (and thus $S\geq 2$) phase diagram under the introduction of
anisotropies, while it is extremely successful both for the isotropic
Heisenberg model and for higher-$S$ spin chains with dimerization? 
In our view the failure is largely and simply due to the fact that the whole
construction of the AKLT model(s) for various spin lengths $S$ resides
entirely on the use of singlet spin-spin bonds. At the isotropic point, each
of these singlet bonds is separated from the neglected triplets by an
energy roughly of the order of the gap energy. Dimerization is
rotationally invariant and accounted for by singlet redistribution between
neighboring bonds. Anisotropies however destroy rotational invariance,
splitting the triplet states, and lower the energetic distance
between the singlet and the lowest triplet state(s). A description of the
chain simply in terms of singlets then quickly becomes inadequate.  

Obviously the $S=1$ phase diagram has a very special property in that
its XY and Haldane phases are separated by $J^{z}=0$ strictly, which
is linked to the fact that the $S=1$ chain can be mapped to a six vertex
times Ising model\cite{Nijs 89}, a mapping which is not possible for
higher spins. The $S=2$ phase diagram reveals no such special property.
It might therefore be conjectured that the $S=1$ case itself is particular
among integer spins and $S=2$ the really generic case.   

\section{Acknowledgements}

All calculations were performed on a PentiumPro 200 MHz Linux and an alpha
533 MHz Linux machine (up to $M=300$, $L=300$), and a Cray T3E
(up to $M=400$, $L=600$). We thank Thierry Jolic\oe ur for useful discussions.

\newpage

%%%%%%%%%%%%%%%%%%%%%%%%%%%%%%%%%%%%%%%%%%%%%%%%%%%%%%%%%%%
\begin{table}
\caption{Position of crossover from alternating to non-alternating
sign behavior of the string order parameter for chain length $L$,
$J^z=1$, and on-site anisotropy $D$.}
\label{tab:table1}
\begin{center}
\begin{tabular}{ccccc}
D & L=50 & L=100 & L=150 & L=200 \\
\hline
0.00 & -- & -- & -- & -- \\
0.01 & -- & -- & -- & -- \\
0.02 & -- & -- & 63 & 59 \\
0.03 & -- & 41 & 39 & 35 \\
0.04 & -- & 29 & 29 & 27 \\
0.05 & 27 & 23 & 23 & 23 \\
0.06 & 21 & 19 & 19 & 19 \\
0.07 & 17 & 17 & 17 & 15 \\
0.08 & 15 & 15 & 15 & 13 \\
0.09 & 13 & 13 & 13 & 13 \\
0.10 & 11 & 11 & 11 & 11 \\
\end{tabular}
\end{center}
\end{table}
%%%%%%%%%%%%%%%%%%%%%%%%%%%%%%%%%%%%%%%%%%%%%%%%%%%%%%%%%%%%

%%%%%%%%%%%%%%%%%%%%%%%%%%%%%%%%%%%%%%%%%%%%%%%%%%%%%%%%%%%
\begin{table}
\caption{Exponent of string order power law decay for various
values of $J^z$ and $D$, extracted from chains of length $L=600$,
keeping $M=400$ DMRG block states.}
\label{tab:table2}
\begin{center}
\begin{tabular}{ccc}
\(\quad J^{z} \quad\) & \( \quad D \quad\) &  \(\quad\alpha\quad \)\\ \hline 
1.0         &  1.5    &    1.24 \\      
1.8         &  1.5    &    1.14 \\      
2.5         &  2.3    &    0.99 \\      
2.7         &  2.4    &    1.00 \\      
2.7         &  2.7    &    1.04 \\      
3.0         &  2.7    &    0.95 \\      
3.3         &  2.8    &    0.86 \\      
3.3         &  3.0    &    0.90 \\      
4.2         &  3.6    &    0.95 \\
\end{tabular}
\end{center}
\end{table}
%%%%%%%%%%%%%%%%%%%%%%%%%%%%%%%%%%%%%%%%%%%%%%%%%%%%%%%%%%%%


\begin{references}  
\bibitem{Haldane 83} F.D.M. Haldane, Phys.\ Lett. {\bf 93A}, 464 (1983),
Phys.\ Rev.\ Lett.\ {\bf 50}, 1153 (1983).
\bibitem{Botet 83} R. Botet, R. Jullien, Phys.\ Rev.\ {\bf B 27}, 613 (1983);
R. Botet, R. Jullien, M. Kolb, Phys.\ Rev.\ {\bf B 28}, 3914 (1983).
\bibitem{White 92} S.R. White, Phys.\ Rev.\ Lett.\ {\bf 69}, 2863 (1992).
\bibitem{White 93a} S.R. White, D.A. Huse, 
Phys.\ Rev.\ {\bf B 48}, 3844 (1993).
\bibitem{Golinelli 94} O. Golinelli, Th.\ Jolic\oe ur, R. Lacaze,
Phys.\ Rev.\ {\bf B 50}, 3037 (1994).
\bibitem{gaps1} J.B. Parkinson and J.C. Bonner, Phys.\ Rev.\ {\bf B 32},
4703 (1985),
M.P. Nightingale and H.W. Bl\"{o}te, Phys.\ Rev.\ {\bf B 33},
650 (1986), A. Moreo, Phys.\ Rev.\ {\bf B 35}, 8562 (1987),
M. Takahashi, Phys.\ Rev.\ {\bf B 38}, 5188 (1988),
K. Nomura, Phys.\ Rev.\ {\bf B 40}, 2421 (1989),
H. Q. Lin, Phys.\ Rev.\ {\bf B 42}, 6561 (1990),
S. Liang, Phys.\ Rev.\ Lett.\ {\bf 64}, 1597 (1990),
K. Kubo, Phys.\ Rev.\ {\bf B 46}, 866 (1992).
\bibitem{Schollwoeck 95} U. Schollw\"{o}ck and Th.\ Jolic\oe ur, Europhys.\
Lett.\ {\bf 30}, 493 (1995); U. Schollw\"{o}ck, O. Golinelli and Th. 
Jolic\oe ur, Phys.Rev. {\bf B 54}, 4038 (1996).
\bibitem{Schollwoeck 96} U. Schollw\"{o}ck and Th.\ Jolic\oe ur, 
Phys.\ Rev.\ Lett.\ {\bf 77}, 2844 (1996).
\bibitem{Yamamoto 95} S. Yamamoto, Phys.\ Rev.\ Lett.\ {\bf 75}, 3348 (1995);
Phys.\ Rev.\ {\bf B 53}, 3364 (1996); Phys.\ Lett.\ {\bf A 213}, 102 (1996).
\bibitem{gaps2} N. Hatano, M. Suzuki, J. Phys.\ Soc.\ Jpn.\ {\bf 62}, 1346
(1993); J. Deisz, M. Jarrell, D.L. Cox, Phys.\ Rev.\ {\bf B 48}, 10227 (1993);
G. Sun, Phys.\ Rev.\ {\bf B 51}, 8370 (1995); Y. Nishiyama, K. Totsuka,
N. Hatano and M. Suzuki, J. Phys.\ Soc.\ Jpn.\ {\bf 64}, 414 (1995);
S. Qin, T.K. Ng, and Z.B. Su, Phys.\ Rev.\ {\bf B 52}, 12844 (1995);
S. Qin, Y.L. Liu, and L. Yu, Phys.\ Rev.\ {\bf B 55}, 2721 (1997).
\bibitem{Golinelli 92b} O. Golinelli, Th.\ Jolic\oe ur, R. Lacaze, 
Phys.\ Rev.\ {\bf B 46}, 10854 (1992). 
\bibitem{Golinelli 93} O. Golinelli, Th.\ Jolic\oe ur, R. Lacaze, 
Phys.\ Rev.\ {\bf B 45}, 9798 (1992). 
\bibitem{Affleck 87} I. Affleck, T. Kennedy, E. Lieb and H. Tasaki, Phys.\
Rev.\ Lett.\ {\bf 59}, 799 (1987); Commun.\
Math.\ Phys.\ {\bf 115}, 477 (1988).
\bibitem{Girvin 89} S.M. Girvin and D.P. Arovas, Phys.\ Scr.\ T {\bf 27}, 
156 (1989).
\bibitem{Kennedy 90} T. Kennedy, J. Phys.: Cond.\ Matt.\ {\bf 2}, 5737 (1990).
\bibitem{Kennedy 92} T. Kennedy and H. Tasaki, Phys.\ Rev.\ {\bf B 45},
304 (1992).
\bibitem{Elstner 94} N. Elstner, H.J. Mikeska, Phys.\ Rev.\ {\bf B 50},
3907 (1994).
\bibitem{Nijs 89} M. den Nijs and K. Rommelse, Phys.\ Rev.\ {\bf B 40}, 
4709 (1989).
\bibitem{Oshikawa 92} M. Oshikawa, J. Phys.\ Cond.\ Matt.\ {\bf 4}, 
7469 (1992).
\bibitem{Schollwoeck 96b} U. Schollw\"{o}ck, Th.\ Jolic\oe ur and Th.\ 
Garel, Phys.\ Rev.\ {\bf B 53}, 3304 (1996).
\bibitem{Schulz 86} H.J. Schulz, Phys.\ Rev.\ {\bf B 34}, 6372 (1986).
\bibitem{Khveshchenko 87} D.V. Khveshchenko, A.V. Chubukov, Sov.\ Phys.\
JETP {\bf 66}, 1088 (1987). 
\bibitem{Oshikawa 95} M. Oshikawa, M. Yamanaka and S. Miyashita, preprint
(cond-mat 9507098)
\bibitem{Renard 87} J.P. Renard, M. Verdaguer, L.P. Regnault, W.A.C. Erkelen.
J. Rossat-Mignod, W.G. Stirling, Europhys.\ Lett.\ {\bf 3}, 945 (1987);
J.P. Renard, L.P. Regnault, M. Verdauguer, J. Phys.\ Coll.\ (Paris),
C8-1425 (1988).
\bibitem{Sakai 90} T. Sakai, M. Takahashi, J. Phys.\ Soc.\ Jpn.\ {\bf 59},
2688 (1990)
\bibitem{Schulz 86b} H.J. Schulz, T. Ziman, Phys.\ Rev.\ {\bf B 33}, 6545
  (1986).
\bibitem{Yajima 94} M. Yajima and M. Takahashi, J. Phys.\ Soc.\ Jpn.\
{\bf 63}, 3634 (1994).
\bibitem{Kitazawa 96} A. Kitazawa, K. Nomura and K. Okamoto, Phys.\ Rev.\ 
Lett.\ {\bf 76}, 4038 (1996).
\bibitem{Papanicolaou 89} N. Papanicolaou, P. Spathis, J. Phys.: Cond.\ Matt.\
{\bf 1}, 5555 (1989); J. Phys.: Cond.\ Matt.\
{\bf 2}, 6575 (1990).
\bibitem{Nomura 97} K. Nomura and A. Kitazawa, preprint (cond-mat 9711294).
\bibitem{Affleck 87b} I. Affleck, F.D.M. Haldane, Phys.\ Rev.\ {\bf B 36},
5291 (1987).
\bibitem{White 93b} S.R. White, Phys.\ Rev.\ {\bf B 48}, 10345 (1993).

\end{references}
\end{document}